\documentclass[12pt]{article}
\pdfoutput=1
\usepackage{latexsym, graphicx,cite}
\usepackage{amsmath}
\usepackage{amssymb}
\usepackage{amsthm}

\usepackage[top = 1. in,bottom = 1. in, left = 1 in, right=1 in]{geometry}

 \newcommand{\arXiv}[1]{\href{http://www.arXiv.org/abs/#1}{arXiv:#1}}
\usepackage[colorlinks=true, linkcolor=blue, bookmarks=true]{hyperref}

\makeatletter
\renewcommand\section{\@startsection {section}{1}{\z@}%
                  {-3.5ex \@plus -1ex \@minus -.2ex}
                  {2.3ex \@plus.2ex}%
                  {\normalfont\large\bfseries}}
\renewcommand\subsection{\@startsection{subsection}{2}{\z@}%
                   {-3.25ex\@plus -1ex \@minus -.2ex}%
                   {1.5ex \@plus .2ex}%
                   {\normalfont\bfseries}}
\makeatother

\newcommand{\beq}{\begin{equation}}
\newcommand{\eeq}{\end{equation}}
\newcommand{\ber}{\begin{array}}
\newcommand{\eer}{\end{array}}
\newcommand{\del}{\partial}
\newcommand{\sssty}{\scriptscriptstyle}

\newcommand{\de}{\delta}

\newcommand{\eps}{\varepsilon}

\newcommand{\om}{\omega}

\newcommand{\ena}{\end{eqnarray}}
\newcommand{\beqa}{\begin{eqnarray}}
\newcommand{\eeqa}{\end{eqnarray}}
\newcommand{\bea}{\begin{eqnarray}}
\newcommand{\eea}{\end{eqnarray}}

\newcommand{\alqu}{\overset{\sssty ?}{\alpha}}

\theoremstyle{remark}

\begin{document}
\begin{titlepage}

\begin{center}
{\LARGE\bf \mbox{Complex plane representations and stationary}\vspace{3mm}\\
states in cubic and quintic resonant systems}\\
\vskip 15mm
{\large Anxo Biasi,$^{a}$ Piotr Bizo\'n$^{\,b}$ and Oleg Evnin$^{c,d}$}
\vskip 7mm
{\em $^a$ Departamento de F\'\i sica de Part\'\i culas, Universidade de Santiago de Compostela
	 and Instituto Galego de F\'\i sica de Altas Enerx\'\i as (IGFAE), Santiago de Compostela, Spain}
\vskip 3mm
{\em $^b$  Institute of Physics, Jagiellonian University, Krak\'ow, Poland}
\vskip 3mm
{\em $^c$ Department of Physics, Faculty of Science, Chulalongkorn University,
Bangkok, Thailand}
\vskip 3mm
{\em $^d$ Theoretische Natuurkunde, Vrije Universiteit Brussel and\\
The International Solvay Institutes, Brussels, Belgium}
\vskip 7mm
{\small\noindent {\tt anxo.biasi@gmail.com, bizon@th.if.uj.edu.pl, oleg.evnin@gmail.com}}
\vskip 10mm
\end{center}
\vspace{1cm}
\begin{center}
{\bf ABSTRACT}\vspace{3mm}
\end{center}

Weakly nonlinear energy transfer between normal modes of strongly resonant PDEs is captured by the corresponding effective resonant systems. In a previous article, we have constructed a large class of such resonant systems (with specific representatives related to the physics of Bose-Einstein condensates and Anti-de Sitter spacetime) that admit special analytic solutions and an extra conserved quantity. Here, we develop and explore a complex plane representation for these systems
modelled on the related cubic Szeg\H o and LLL equations. To demonstrate the power of this representation, we use it to give simple closed form expressions for families of stationary states bifurcating from all individual modes. The conservation laws, the complex plane representation and the stationary states admit furthermore a natural generalization from cubic to quintic nonlinearity. We demonstrate how two concrete quintic PDEs of mathematical physics fit into this framework, and thus directly benefit from the analytic structures we present: the quintic nonlinear Schr\"odinger equation in a one-dimensional harmonic trap, studied in relation to Bose-Einstein condensates, and the quintic conformally invariant wave equation on a two-sphere, which is of interest for AdS/CFT-correspondence.

\vfill

\end{titlepage}


\section{Introduction}

Resonant Hamiltonian systems of the form
\beq
i\,\frac{d\alpha_n}{dt}=\hspace{-2mm}\sum_{n+m=k+l} \hspace{-2mm}C_{nmkl}\bar\alpha_m\alpha_k\alpha_l
\label{ressyst}
\eeq
emerge in weakly nonlinear analysis of strongly resonant PDEs, with many applications in mathematical physics. Here, $\alpha_n(t)$ with an integer $n\ge 0$ are complex-valued dynamical variables, the bar denotes complex conjugation, and $C_{nmkl}$ are numbers known as the mode couplings or the interactions coefficients. Note that the summation is restricted by the \emph{resonant constraint} $n+m=k+l$. To quickly demonstrate how equations of this sort originate in a physically motivated example, consider the simplest possible setting afforded by the one-dimensional cubic nonlinear Schr\"odinger equation on the real line in a harmonic potential
\begin{equation}
i\,\frac{\partial \Psi}{\partial t}=\frac12\left(-\frac{\partial^2}{\partial x^2}+x^2\right)\Psi +g|\Psi|^2\Psi.
\label{NLS1d}
\end{equation}
At $g=0$, one has the linear Schr\"odinger equation of a harmonic oscillator, whose general solution is
\begin{equation}
\Psi=\sum_{n=0}^\infty \alpha_n \psi_n(x) e^{-iE_n t},
\label{NLS1dlin}
\end{equation}
with constant $\alpha_n$, where 
\beq
\qquad E_n=n+\frac12
\eeq
are the eigenstate energies, and $\psi_n$ are the corresponding normalized wavefunctions satisfying
\beq
 \frac12\left(-\frac{\partial^2}{\partial x^2}+x^2\right)\psi_n=E_n\psi_n.
\eeq
 At a small nonzero coupling $g$, $\alpha_n$ are no longer constant and evolve slowly over time. This evolution can be extracted by substituting \eqref{NLS1dlin} into \eqref{NLS1d} and projecting on $\psi_k(x)$, which gives
\begin{equation} 
 i\,\frac{d\alpha_n (t)  }{dt}= g\sum_{k,l,m=0}^\infty C_{nmkl} \,\bar \alpha_m  (t)\alpha_k (t) \alpha_l (t)\,e^{i(E_n+E_m-E_k-E_l)t}, 
\label{NLSpreres}
\end{equation}
with $C_{nmkl}=\int dx \,\psi_n  \psi_m \psi_k \psi_l$. At $g\ll 1$, one makes use of the \emph{resonant approximation}, which consists in dropping all oscillatory terms on the right-hand side, keeping only those satisfying $E_n+E_m-E_k-E_l=0$, which is the same as $n+m=k+l$. This is known to result in an accurate approximation of the full equation at small $g$ on very long time scales of order $1/g$.
This approximation, known by a number of different names, goes back to the foundational work of Bogoliubov and Krylov \cite{BK} on `time-averaging.' Pedagogical introductions can be found in \cite{BM,murdock, KM}, while a variety of applications in contemporary literature can be sampled from \cite{bambusi,ZhP,ZhPmore,KSS,FGH,GHT,GT}. Once the resonant approximation has been implemented, and $g$ has been absorbed in a redefinition of time, one ends up with (\ref{ressyst}).

While we have chosen the simple equation (\ref{NLS1d}) to illustrate the underlying idea, the resonant approximation leads to equations of the form (\ref{ressyst}) in a number of other interesting cases, of course, with different assignments of the interaction coefficients $C_{nmkl}$. We mention here the applications to higher-dimensional nonlinear Schr\"odinger equations in harmonic traps \cite{GHT,GT,BMP,BBCE,GGT}, studies of nonlinear wave equations on spheres and in Anti-de Sitter spacetimes \cite{CF,BHP,BEL} as well as studies of weakly nonlinear gravitational dynamics in Anti-de Sitter spacetimes \cite{FPU,CEV,BMR,islands,AdS4,BEF} in relation to its nonlinear instability \cite{BR,rev2}. A resonant equation of the form (\ref{ressyst}) has recently been recovered as well for an asymptotically Anti-de Sitter wormhole spacetime \cite{resscalar1,resscalar2}, while a general algorithm exists \cite{KG} for constructing spacetimes with resonant structures underlying (\ref{ressyst}). The particular case $C_{nmkl}=1$, known as the cubic Szeg\H o equation, is Lax-integrable and has been studied as an integrable model of turbulence in the mathematical literature \cite{GG} with powerful results. (We note in passing that perhaps the most familiar spatially confined setting where a translationally invariant PDE is compactified on a torus does not possess the type of resonant spectrum of linearized perturbation frequencies underlying our studies.)

Many of the cases listed above generate resonant systems possessing a rich algebraic structure, including special analytic solutions in the fully nonlinear regime and extra conserved quantities \cite{GHT,BBCE,CF,BHP,BEL, BEF}. This has led us to formulating conditions on the interaction coefficients $C_{nmkl}$ that guarantee such special properties, resulting in a large class of partially solvable resonant systems presented in \cite{AO}. Our purpose in this paper is to further explore the properties of this class of systems, and to construct its generalizations. To this end, we start by devising a complex plane representation for the partially solvable resonant systems of \cite{AO} formulated as an integro-differential equation for the generating function of $\alpha_n$. Such representations have proven very effective, for example, for the cubic Szeg\H o equation \cite{GG} and the LLL equation \cite{GHT,GT}, the latter equation being a member of our partially solvable class.\footnote{We owe some inspiration to Patrick G\'erard who has shown to us that the complex plane representation of the conformal flow developed in \cite{CF} can be used to give a lightning-speed proof of the known fact that all Blaschke product functions give rise to stationary states of the conformal flow (which is, again, a particular representative of the partially solvable class of \cite{AO}). Similarly, powerful results for the asymptotics of stationary states at spatial infinity and the distribution of their zeros are derived for the LLL equation in \cite{GGT} relying on a complex plane representation.} After reviewing the material of \cite{AO} and constructing the complex plane representation in section 2, we use this representation, in section 3, to prove a simple closed-form formula for stationary solutions bifurcating from individual modes.

In section 4, we present a quintic generalization of the partially solvable resonant systems of \cite{AO}, and of the new material of sections 2 and 3. The reason to look for such generalization is twofold. First, it is interesting to see which of the partially solvable features seen for the cubic equation (\ref{ressyst}) are structurally stable and survive a quintic generalization (for instance, there is no known quintic generalization \cite{gerard_quintic} of the solvable structures of the cubic Szeg\H o equation, a Lax-integrable resonant system outside the partially solvable class of \cite{AO}). Second, in physically motivated situations, one often has the field value reflection symmetry that only allows odd powers of the field in the equations of motion. Under such circumstances, if the cubic term is tuned down to zero, the quintic one becomes the dominant nonlinearity that governs the weakly nonlinear dynamics. This is literally possible in trapped Bose-Einstein condensates, where the cubic term can be switched off using Feshbach resonances, leading to the emergence of a quintic nonlinear Schr\"odinger equation \cite{NLSquint}. Finally, in section 5, we present two concrete quintic PDEs of mathematical physics whose resonant systems fit into our setup: first, the quintic nonlinear Schr\"odinger equation in a one-dimensional harmonic trap (recently treated in \cite{fennell}), and second, the quintic conformally invariant wave equation on a two sphere (which generalizes the cubic considerations on a three-sphere in \cite{CF}).


\section{Complex plane representation for cubic resonant systems}\label{sec:cpl}

We start with briefly reviewing the main points of \cite{AO}. 
One defines a large class of resonant systems of the form (\ref{ressyst}) by specifying conditions on the interaction coefficients $C_{nmkl}$ that result in far-reaching analytic properties. To this end, we first introduce a positive real number $G$ (to be specified for each individual resonant system in our class) and define
\beq
f_n=\sqrt{\frac{(G)_n}{n!}},
\label{fdef}
\eeq
where $(G)_n\equiv G(G+1)\cdots(G+n-1)$ is the Pochhammer symbol. We furthermore define
\beq
\beta_n=\frac{\alpha_n}{f_n},\qquad S_{nmkl}=f_nf_mf_kf_lC_{nmkl}.
\label{betaSdef}
\eeq
The key condition of \cite{AO} we shall impose on $C_{nmkl}$ is conveniently stated in terms of $S_{nmkl}$ as
\beq
\left(n -1+ G\right)S_{n-1,mkl} + \left(m - 1 + G\right)S_{n,m-1,kl}
- (k+1)S_{nm,k+1,l} - (l+1)S_{nmk,l+1} =0.
\label{AOdef}
\eeq
If this condition is fulfilled\footnote{We adopt the convention, here and for the rest of the paper, that if any of the indices is negative, the corresponding value of $S$ is zero.} for some specific number $G$ and for all mode number quartets $(n,m,k,l)$ satisfying $n+m-1=k+l$, the following properties of the corresponding resonant system (\ref{ressyst}) are ensured:
\begin{itemize}
\item The resonant system (\ref{ressyst}) respects the conservation of
\beq
Z  =  \sum_{n=0}^{\infty}\sqrt{(n+1)(n + G)}\,\bar{\alpha}_{n+1} \alpha_n.
\label{Zdef}
\eeq
\item There is an invariant manifold of the evolution defined by (\ref{ressyst}) given by
\beq
\beta_n = \big( b(t)+n\, a(t)\big) (p(t))^n,
\eeq
where $a(t),b(t),p(t)$ are complex-valued functions of time.
\item Within this invariant manifold, the dynamics is Liouville-integrable, and the spectrum $|\beta_n(t)|^2$ is exactly periodic in time.
\end{itemize}
Such features were first seen in concrete resonant systems arising from PDEs of mathematical physics \cite{CF,BEL,BBCE}, which led to the formulation of the general condition (\ref{AOdef}) from which they follow. We note that there is a special case that can be seen as a $G\to\infty$ limit, explicitly described in \cite{AO}. This special case amounts to typographically replacing all expressions of the form ``(integer+$G$)'' in the above formulas by 1, and correspondingly replacing $(G)_n$ in (\ref{fdef}) by 1. The resulting structure remains valid, and is in fact physically realized in the Landau level truncations of the resonant approximation to the Gross-Pitaevskii equation \cite{BBCE}. The details can be found in \cite{AO}. For brevity, we shall not be treating  this special case here, and will concentrate on generic values of $G$.

The finite difference equation (\ref{AOdef}) can be resolved in terms of the generating function for $S_{nmkl}$ and, as shown in  \cite{AO}, is equivalent to stating that
\beq\label{S_sol}
S(y,z,v,w)=\sum_{n,m,k,l} S_{nmkl}\, y^nz^mv^kw^l=\frac{{\cal F}\left(\ln\left[\frac{(1-vy)(1-wz)}{(1-vz)(1-wy)}\right]\right)}{\left[(1-vy)(1-vz)(1-wy)(1-wz)\right]^{G/2}},
\eeq
where $\cal F$ is an arbitrary even function, ${\cal F}(x)={\cal F}(-x)$. One can set ${\cal F}(0)=1$ by rescaling the time variable. We note (and this shall be used below) that since the generating function depends only on $vy$, $vz$, $wy$ and $wz$, its power series expansion contains only terms with $n+m=k+l$. Thus summing over all $n,m,k,l$ in the definition of the generating function is completely equivalent to only summing over the resonant quartets $n+m=k+l$.

The resonant system (\ref{ressyst}) can be expressed through $\beta_n$ defined by (\ref{betaSdef}) as
\beq
i\frac{(G)_n}{n!}\frac{d\beta_n}{dt}=\hspace{-2mm}\sum_{n+m=k+l} \hspace{-2mm}S_{nmkl}\bar\beta_m\beta_k\beta_l.
\label{betares}
\eeq
In order to take advantage of (\ref{S_sol}), we introduce the following generating functions for $\beta_n$:
\beq
u(t,z)=\sum_{n=0}^\infty \beta_n z^n,\qquad \tilde u(t,z)=\sum_{n=0}^\infty\frac{\bar\beta_n}{z^n},
\label{udef}
\eeq
so that
\beq
\beta_n(t)=\frac1{2\pi i}\oint \frac{dz }{z^{n+1}}u(z,t),\qquad \bar\beta_n=\frac1{2\pi i}\oint dz \,z^{n-1}\,\tilde u(z,t) .
\label{betaint}
\eeq
Note that the integration contour for $\beta_n$ must not enclose any singularities of $u$, while the integration contour for $\bar\beta_n$ must enclose all singularities of $\tilde u$. The tilde-conjugation can be understood as taking complex conjugates of the values of $u$ on the unit circle, and then analytically continuing away from the unit circle to obtain $\tilde u$. One can also write
\beq
\tilde u(z)=\overline{u\left(1/\bar z\right)}.
\label{tildeconj}
\eeq

Substituting (\ref{betaint}) to (\ref{betares}), multiplying with $z^n$ and summing over $n$, one gets
\beq
\frac{i}{\Gamma(G)}\del_t\del_z^{G-1}(z^{G-1} u(t,z))=\frac1{(2\pi i)^3}\oint\frac{ds}s \oint\frac{dv}v \oint\frac{dw}w S(z,s,1/v,1/w)\tilde u(t,s) u(t,v) u(t,w),
\label{complexeq}
\eeq
where we have used the fact that the constraint $n+m=k+l$ in the summation can be ignored since $S_{nmkl}$ extracted from the generating function $S$ automatically vanish unless this constraint is satisfied. The integration contours for $v$ and $w$ are outside the  unit circle but do not enclose any singularities of $u$, while the integration contour for $s$ is inside the unit circle and encloses all singularities of $\tilde u$. Note that to ensure convergence in the resummation of $S$ according to (\ref{S_sol}), $|v|$ and $|w|$ must be greater than both $|z|$ and $|s|$. The fractional derivative $\del_z^{(G-1)}$ is defined by its action on powers of $z$
\beq
\del_z^a z^b=\frac{\Gamma(b+1)}{\Gamma(b-a+1)}z^{b-a},
\label{fracdef}
\eeq
with $\Gamma$ being the usual Euler's $\Gamma$-function.
This simple-minded definition goes back to the very origins of the fractional calculus \cite{frcalc1,frcalc2}. In general, acting with fractional derivatives on integer powers of $z$ produces fractional powers, which are not single-valued, which is a source of ambiguities in defining fractional calculus. Note, however, that applying $\del_z^a z^a$ to an analytic function, which is the only operation we need for (\ref{complexeq}), always results in an analytic function, hence no subtleties occur in this case. The way (\ref{fracdef}) enters the complex plane representation of (\ref{betaint}) is through the relation $\Gamma(n+G)/\Gamma(n+1)=\Gamma(G)\,\,(G)_n/n!$.

We note that the differentiation defined by (\ref{fracdef}) can be equally well implemented via a Cauchy-like complex contour formula for the operator $\del_z^a z^a$ featured in (\ref{complexeq}) acting on a holomorphic function $f(z)$:
\beq
\del_z^a (z^a f(z))=\frac{\Gamma(a+1)}{2\pi i}\oint \frac{s^a\, f(s)\,ds}{(s-z)^{a+1}}.
\label{fracCauchy}
\eeq
Indeed, expanding $f(s)$ in terms of integer powers of $s$, we see that (\ref{fracCauchy}) acts on the individual powers in accordance with (\ref{fracdef}), as can be verified by evaluating the residue at $s=\infty$. (Note that, for noninteger $a$, there is a cut in the complex plane, but it only extends from $s=0$ to $s=z$, without affecting the evaluation of the residue at $s=\infty$. The integration contour is defined to lie outside this cut.)

A representation of the form (\ref{complexeq}) has been previously obtained for the conformal flow \cite{CF}, which corresponds in our present language to $G=2$ (while the right-hand side integral can be further simplified due to particular factorization properties of $S$ in the case of the conformal flow). Our present derivation has established this representation for the entire class of partially solvable resonant systems of \cite{AO}, of which the conformal flow is a representative.
One can substitute (\ref{S_sol}) into (\ref{complexeq}) to obtain explicitly
\beq
\frac{i}{\Gamma(G)}\del_t\del_z^{G-1}(z^{G-1} u(t,z))=\frac1{(2\pi i)^3}\oint\frac{ds\,dv\,dw}{s\,(vw)^{1-G}} \frac{\tilde u(t,s) u(t,v) u(t,w) \,\,{\cal F}\left(\ln\left[\frac{(v-s)(w-z)}{(v-z)(w-s)}\right]\right)}{\left[(v-s)(v-z)(w-s)(w-z)\right]^{G/2}},
\label{complexeq2}
\eeq
Note the emergence of the cross-ratio $(v-s)(w-z)/(v-z)(w-s)$, which is a conformally invariant combination of the coordinates of four points $(z,s,v,w)$ on the complex plane \cite{appconf}.


\section{Stationary states}\label{sec:stst}

We shall consider stationary states of the form
\beq
\alpha_n= e^{-i\lambda t} A_n
\eeq
that solve (\ref{ressyst}), where $\lambda$ is a real number and $A_n$ are constants. Evidently, because of the resonant structure of (\ref{ressyst}),
\beq
A_n=0\qquad\mbox{if}\quad n\ne k,
\label{singlemode}
\eeq
provide such solutions for any given $k$. These single-mode solutions supported by mode number $k$ are present in any resonant system of the form (\ref{ressyst}), irrespectively of the values of the interaction coefficients $C_{nmkl}$.

For the partially solvable resonant systems of \cite{AO} that possess the structures outlined in the previous section, it is possible to go further and give a simple explicit formula for inifinite families of stationary solutions bifurcating from each single-mode solution. This section is dedicated to developing these formulas. One particular purpose is to demonstrate the power of the complex plane representation (\ref{complexeq2}) through the way it facilitates the analysis of these stationary states. The existence of the families bifurcating from the solutions (\ref{singlemode}) may be anticipated from the presence of the symmetry generated by the conserved quantity (\ref{Zdef}), which may act on the states (\ref{singlemode}) to produce new solutions. We are not aware, however, of a straightforward way to integrate the infinitesimal transformations generated by (\ref{Zdef}) to finite transformations on the infinite-dimensional configuration space parametrized by $\alpha_n$, and resort to direct analysis of the equations of motion in the form (\ref{complexeq2}).

Below, we start by presenting the stationary state bifurcating from mode 0, for which the analysis is particularly simple, and allows for a transparent demonstration of the underlying idea. We then proceed with the general consideration for stationary states bifurcating from higher modes.

\subsection{Stationary states bifurcating from the lowest mode}

We claim that
\beq
u(t,z)=\frac{e^{-i\lambda t}}{1-pz}
\label{mode0fam}
\eeq
solves (\ref{complexeq}) for any complex value of  $p$ (and some $p$-dependent value of $\lambda$). Note that if one sends $p$ to 0, one obtains $u(t,z)=e^{-i\lambda t}$, which does not depend on $z$ and corresponds to $\alpha_{n\ge0}=0$, i.e., to the single-mode stationary states supported by mode 0. 
Thus, our family (\ref{mode0fam}) bifurcates from mode 0, as anticipated in the title.

The l.h.s.\ of (\ref{complexeq2}) becomes simply 
\beq
\frac{\lambda e^{-i\lambda t}}{(1-pz)^G},
\label{mode0lhs}
\eeq
as one can easily see by applying (\ref{fracCauchy}) and evaluating the residue at $1/p$.
For the r.h.s.\ of (\ref{complexeq2}), we first note that
\beq
\tilde u(t,z)=\frac{e^{i\lambda t}}{1-\bar p/z}.
\label{mode0famtilde}
\eeq
Hence, the r.h.s.\ of (\ref{complexeq2}) may be  written as
\beq
\frac{e^{-i\lambda t}}{(2\pi i)^3}\oint\frac{ds\,dv\,dw}{s\,(vw)^{1-G}} \frac{{\cal F}\left(\ln\left[\frac{(v-s)(w-z)}{(v-z)(w-s)}\right]\right)}{\left[(v-s)(v-z)(w-s)(w-z)\right]^{G/2}}\frac{1}{1-\bar p/s} \frac{1}{1-pv} \frac{1}{1-pw},
\eeq
Consider first the integral over $v$. There are branch cuts connecting $v=0$, $v=z$ and $v=s$, but all of these branch cuts are inside the integration contour, while the simple pole at $v=1/p$ is outside the contour. Thus, the integral can be evaluated as the residue at $v=1/p$. The same argument applies to the integral over $w$. Implementing these two operations, one gets:
\beq
\frac1{(1-pz)^G}\frac{e^{-i\lambda t}}{2\pi i}\oint \frac{ds}{(s-\bar p)(1-sp)^G}.
\eeq
Note that at $v=w=1/p$ the argument of ${\cal F}$ has turned into 0, while we have assumed ${\cal F}(0)=1$ by a choice of the time scale. The function ${\cal F}$ has thus dropped out from our expression at this stage. As far as the remaining integral over $s$ is concerned, 
once again, there is a branch cut outside the integration contour, but inside there is only a simple pole, so one can express the result through the residue, obtaining
\beq
\frac{1}{(1-|p|^2)^G}\frac{e^{-i\lambda t}}{(1-pz)^G}.
\eeq
This expression for the r.h.s.\ of (\ref{complexeq2}) manifestly matches the l.h.s.\ given by (\ref{mode0lhs}).
The equations of motion are thus satisfied by our family of stationary states (\ref{mode0fam}) provided that
\beq
\lambda=\frac{1}{(1-|p|^2)^G}.
\eeq
Note that this holds for every value of $G$, and irrespectively of the form of the arbitrary function $\cal F$ contained in the definition (\ref{S_sol}) of our class of resonant systems.

\subsection{Stationary states bifurcating from higher modes}

We now proceed with the stationary solutions bifurcating from mode number $N$. We claim that the relevant generating function $u(t,z)$ satisfies
\beq
u(t,z)=e^{-i\lambda t}u(z),\qquad\del_z^{G-1}(z^{G-1} u(z))=\frac{(\bar p -z)^N}{(1-pz)^{N+G}}.
\label{uNder}
\eeq
Indeed, if $p=0$, $\del_z^{G-1}(z^{G-1} u)$ is proportional to $z^N$, and hence $u$ itself is proportional to $z^N$, i.e., the only nonvanishing $\alpha_n$ is $\alpha_N$. (The above formula, as well as (\ref{mode0fam}), originated as a guess based on numerical experimentation, before being provided an analytic proof that we are about to present.)

We have specified $u(z)$ through the result of acting on it with $\del_z^{G-1}z^{G-1}$. What about $u(z)$ itself? We can say that it is of the form
\beq
u(z)=\sum_{k=0}^{N} \frac{c_k}{(1-pz)^{k+1}},
\label{formuN}
\eeq
though we are not aware of simple explicit expressions for $c_k$. Differentiation of the individual terms in (\ref{formuN}) follows the rule
\beq
\del_z^{G-1}\frac{z^{G-1} }{(1-pz)^{k+1}}=\frac{\Gamma(G)}{2\pi i}\oint \frac{ds}{(s-z)^G}\frac{s^{G-1} }{(1-ps)^{k+1}}=\frac{\Gamma(G)}{(-p)^{k+1}}\,\del^{k}_s\frac{s^{G-1} }{(s-z)^G}\Bigg|_{s=1/p}.
\eeq
The last expression is evidently a linear combination of terms of the form $1/(1-pz)^G$, $1/(1-pz)^{G+1}$, ..., $1/(1-pz)^{G+k}$. Hence, $\del_z^{G-1}(z^{G-1} u(t,z))$ is a linear combinations of terms of the form $1/(1-pz)^G$, $1/(1-pz)^{G+1}$, ..., $1/(1-pz)^{G+N}$ with coefficients that are themselves linear combinations of $c_0$, $c_1$, ..., $c_{N}$. By tuning these $N+1$ coefficients, we can make $\del_z^{G-1}(z^{G-1} u(t,z))$ equal $1/(1-pz)^{G+N}$ times an arbitrary polynomial of degree $N$ in $z$, and in particular, we can make it equal (\ref{uNder}).

With these preliminaries, the l.h.s.\ of (\ref{complexeq}) is by construction
\beq
\frac{\lambda\,e^{-i\lambda t}}{\Gamma(G)}\frac{(\bar p -z)^N}{(1-pz)^{N+G}},
\label{lhs}
\eeq
and we have to prove that the r.h.s.\ of (\ref{complexeq}), given by 
\beq
\frac{e^{-i\lambda t}}{(2\pi i)^3}\oint\frac{ds\,dv\,dw}{s\,(vw)^{1-G}} \frac{\tilde u(s) u(v) u(w) \,\,{\cal F}\left(\ln\left[\frac{(v-s)(w-z)}{(v-z)(w-s)}\right]\right)}{\left[(v-s)(v-z)(w-s)(w-z)\right]^{G/2}},
\label{rhsuuu}
\eeq
matches this form. Our proof will proceed in two steps (which can be thought of as lemmas). In step 1, we shall show that (\ref{rhsuuu}) must be of the form
\beq
\frac{Q_N(z)\,e^{-i\lambda t}}{(1-pz)^{N+G}},
\label{rhsQ}
\eeq
where $Q_N(z)$ is a polynomial of degree $N$ in $z$. In step 2, we shall show that (\ref{rhsuuu}) and its first $N-1$ $z$-derivatives must vanish at $z=\bar p$, which means that  (\ref{rhsuuu}) has a degree $N$ zero at that point. Combined, these two facts imply that (\ref{rhsuuu}) is proportional to 
\beq
e^{-i\lambda t}\frac{(\bar p -z)^N}{(1-pz)^{N+G}},
\label{rhsfnl}
\eeq
The coefficient of proportionality simply fixes $\lambda$, and in view of (\ref{lhs}), the equation of motion (\ref{complexeq2}) is satisfied. We now proceed filling in the details of step 1 and step 2 required to complete the proof. In handling (\ref{rhsuuu}) below, we shall suppress the factor $e^{-i\lambda t}$ which is common to the entire expression (\ref{rhsuuu}) and already matches (\ref{lhs}).\vspace{2mm}

\noindent{\bf Step 1.} With (\ref{formuN}) in mind, the integrals over $v$ and $w$ in (\ref{rhsuuu}) can be evaluated in terms of residues as a linear combination of terms of the form 
\beq
\frac1{2\pi i}\oint\frac{ds}s \tilde u(s)\,\,\del_v^k\del_w^l\Bigg(\frac{{\cal F}\left(\ln\left[\frac{(v-s)(w-z)}{(v-z)(w-s)}\right]\right)}{(vw)^{1-G}\left[(v-s)(v-z)(w-s)(w-z)\right]^{G/2}}\Bigg)\Bigg|_{v,w=1/p}
\label{rhskl}
\eeq
with
\beq
0\le k,l\le N.
\label{klN}
\eeq
Now, for any $\cal G$ that depends on the indicated argument,
\beq
\del_v{\cal G}\left(\ln\left[\textstyle\frac{(v-s)(w-z)}{(v-z)(w-s)}\right]\right)=\left(\frac{1}{v-s}-\frac{1}{v-z}\right){\cal G\,}'\left(\ln\left[\textstyle\frac{(v-s)(w-z)}{(v-z)(w-s)}\right]\right).
\label{diffG}
\eeq
Applying such differentiations recursively, each $v$-derivative may produce either one extra factor of $1/(v-z)$ if it acts on the numerator in (\ref{rhskl}) or on the factor $1/(v-z)^{G/2}$, or it may produce one extra factor of $1/(v-s)$ in a similar manner, or it may produce simply extra factors of $v$ if it acts on $1/v^{1-G}$ (such factors will be expressed through $p$ only after one substitutes $v=1/p$ at the end, and are irrelevant for our present argument). Furthermore, in the process of differentiation, $\cal F$ may change into another function of the same argument. The situation with $w$-differentiations is, of course, directly parallel. We denote the number of extra factors of $1/(v-z)$ generated through such differentiations as $k_1$, the number of extra factors $1/(v-s)$ as $k_2$, the number of extra factors of $1/(w-z)$ as $l_1$, and the number of factors $1/(w-s)$ as $l_2$. Evidently, from the above description of differentiations, 
\beq
k_1+k_2\le k\quad\mbox{and}\quad l_1+l_2\le l.
\label{k1k2k}
\eeq
 One concludes that $\del_v^k\del_w^l(\ldots)$ in (\ref{rhskl}) consists of terms of the  form
\beq
\frac{{\cal G}\left(\ln\left[\textstyle\frac{(v-s)(w-z)}{(v-z)(w-s)}\right]\right)}{(v-z)^{k_1+G/2}(v-s)^{k_2+G/2}(w-z)^{l_1+G/2}(w-s)^{l_2+G/2}},
\eeq
where we have omitted the $(s,z)$-independent prefactor, and $\cal G$ is some function expressed through $\cal F$ and its derivatives. Once $v=w=1/p$ have been substituted, the argument of $\cal G$ turns into 0, so it is just a number. At the end, once again ignoring $(s,z)$-independent factors, (\ref{rhskl}) is written as a linear combination of terms of the form
\beq
\frac{1}{(1-pz)^{G+k_1+l_1}}\frac1{2\pi i}\oint  \frac{ \tilde u(s)\,ds}{s\,(1-ps)^{G+k_2+l_2}}.
\label{k1k2l1l2}
\eeq
Now, with $\sigma=1/s$,
\beq
\frac1{2\pi i}\oint ds \frac{\tilde u(s)}{s\,(1-ps)^{G+k_2+l_2}}=\frac1{2\pi i}\oint d\sigma\frac{\sigma^{G+k_2+l_2-1} \tilde u(1/\sigma)}{(\sigma-p)^{G+k_2+l_2}}.
\eeq
But $\sigma^{k_2+l_2}/(\sigma-p)^{k_2+l_2}$ can be written as a linear combination of terms of the form $ 1/(\sigma-p)^{m}$ with $0\le m\le k_2+l_2$. Hence, (\ref{k1k2l1l2}) can be written as a linear combination  of
\beq
\frac1{2\pi i}\oint d\sigma\frac{\sigma^{G-1} \tilde u(1/\sigma)}{(\sigma-p)^{G+m}}=\frac1{\Gamma(G+m)}\del_\sigma^{G+m-1}\left(\sigma^{G-1} \tilde u(1/\sigma)\right)\Big|_{\sigma=p}.
\label{usigmares}
\eeq
By (\ref{tildeconj}), $\tilde u(1/\sigma)$ is exactly the same as $u(z)$ with $z$ replaced by $\sigma$ and $p$ replaced by $\bar p$. Therefore, from (\ref{uNder}),
\beq
\del_\sigma^{G-1}\left(\sigma^{G-1}\tilde u(1/\sigma)\right)=\frac{(p -\sigma)^N}{(1-\bar p\sigma)^{N+G}},
\eeq
which evidently has a degree $N$ zero at $\sigma=p$. Hence, at least $N$ further differentiations ($m \ge N$) must be applied in (\ref{usigmares}) in order for the result to be nonzero. So (\ref{k1k2l1l2}) can only be nonvanishing if $k_2+l_2\ge N$. But by (\ref{klN}) and (\ref{k1k2k}) this implies $k_1+l_1\le N$. Thus, after evaluating of the $s$-integral in (\ref{k1k2l1l2}), the result is a linear combination of $1/(1-pz)^{G+n}$ with $0\le n\le N$. Since all contributions to (\ref{rhsuuu}) are in the form (\ref{k1k2l1l2}), this property is inherited by (\ref{rhsuuu}), and hence the latter must then be expressible as (\ref{rhsQ}).\vspace{2mm}

\noindent{\bf Step 2.} To pin down $Q_N(z)$ in (\ref{rhsQ}), we shall compute the $k$th $z$-derivative of (\ref{rhsuuu}) at $z=\bar p$, and we shall start with performing the $s$-integral in (\ref{rhsuuu}). The structure of the argument is rather similar to step 1, but with the roles of $(z,s)$ and $(v,w)$ interchanged. The $s$-integral is evaluated through the residues of $\tilde u$. The following expression for $\tilde u$ follows from (\ref{formuN}): 
\beq
\tilde u(s)=\sum_{l=0}^{N} \frac{\bar c_l}{(1-\bar p/s)^{l+1}}=s\sum_{l=0}^{N} \frac{d_l}{(s-\bar p)^{l+1}},
\label{formutild}
\eeq
where $d_l$ are certain linear combinations of $\bar c_l$ with $\bar p$-dependent coefficients.\footnote{The explicit expression for $d_l$ is irrelevant for the purposes of our argument. For any fixed $N$, it is easily derived by multiplying the numerator and denominator of ${\bar c_l}/(1-\bar p/s)^{l+1}$ by $s^{l+1}$ and then expanding ${s^l}/(s-\bar p)^{l+1}$ in terms of $1/(s-\bar p)$, $1/(s-\bar p)^2$,... , $1/(s-\bar p)^{l+1}$. For instance, at $N=1$, $d_0=\bar c_0+\bar c_1$ and $d_1=\bar p\bar c_1$.} Then the $k$th $z$-derivative of (\ref{rhsuuu}) at $z=\bar p$ consists of terms of the form
\beq
\frac1{(2\pi i)^2}\oint\frac{dv\,dw}{(vw)^{1-G}} \,u(v) u(w)\del_z^k\del_s^l\Bigg( \frac{{\cal F}\left(\ln\left[\frac{(v-s)(w-z)}{(v-z)(w-s)}\right]\right)}{\left[(v-s)(v-z)(w-s)(w-z)\right]^{G/2}}\Bigg)\Bigg|_{z,s=\bar p} .
\eeq
The $z$- and $s$-differentiations are performed in a manner directly parallel to the $v$- and $w$- differentiations in step 1. One gets a collection of terms of the form
\beq
\frac1{(2\pi i)^2}\oint\frac{dv}v \oint\frac{dw}w  \frac{u(v) u(w)}{v^{k_1+l_1}w^{k_2+l_2}(1-\bar p/v)^{G+k_1+l_1}(1-\bar p/w)^{G+k_2+l_2}},
\eeq
with $k_1+k_2=k$ and $l_1+l_2=l$, which can be rewritten as
\beq
\del_v^{G-1+k_1+l_1}(v^{G-1}u(v))\, \del_w^{G-1+k_2+l_2}(w^{G-1}u(w))\Big|_{v,w=\bar p}.
\eeq
But by construction $\del_z^{G-1}(z^{G-1}u(z))$ has a degree $N$ zero at $z=\bar p$. Therefore, at least $N$ extra differentiations must be applied in both factors above in order to make the result nonvanishing, i.e., $k_1+l_1\ge N$ and $k_2+l_2\ge N$, while $l_1+l_2=l\le N$ from (\ref{formutild}), and hence $k=k_1+k_2\ge N$. In other words, if we apply fewer than $N$ $z$-derivatives to (\ref{rhsuuu}) and evaluate the result at $z=\bar p$, all the terms identially vanish. That means that (\ref{rhsuuu}) has a degree $N$ zero at $z=\bar p$. Since we already know that it is of the form (\ref{rhsQ}), it must be proportional to (\ref{rhsfnl}), completing our proof that (\ref{uNder}) satisfies the equations of motion (\ref{complexeq2}). Note that the specific form of the arbitrary function $\cal F$ defining our resonant system does not affect the form of stationary solutions (\ref{uNder}), though it may affect the relation of $\lambda$ and $p$.


\section{A quintic generalization}

In this section, we address the question which of the properties of cubic resonant systems we have previously described generalize to the  quintic analog of (\ref{ressyst}) given by
\beq
i\,\frac{d\alpha_n}{dt}=\hspace{-4mm}\sum_{n+n_2+n_3=k_1+k_2+k_3} \hspace{-4mm}C_{nn_2n_3k_1k_2k_3}\bar\alpha_{n_2}\bar\alpha_{n_3}\alpha_{k_1}\alpha_{k_2}\alpha_{k_3}.
\label{res5}
\eeq
We shall see that most (but not all) of the properties discussed above have a natural generalization, provided that a suitable constraint is imposed on $C$. This is in contrast, for example, to the cubic Szeg\H o equation whose integrability properties have no known quintic generalization \cite{gerard_quintic}. 

We start by asking whether the conservation of $Z$ given by (\ref{Zdef}) can be ensured for some value of $G$. By analogy with (\ref{betaSdef}), define
\beq
S_{n_1n_2n_3k_1k_2k_3}=f_{n_1}f_{n_2}f_{n_3}f_{k_1}f_{k_2}f_{k_3}C_{n_1n_2n_3k_1k_2k_3}
\label{Squintdef}
\eeq
with $f_n$ given by (\ref{fdef}). Note that $C$, and hence $S$, are symmetric under any permutations of $(n_1,n_2,n_3)$, any permutations of $(k_1,k_2,k_3)$ and under interchange of these two groups of indices. One can then check that (\ref{Zdef}) is indeed conserved by (\ref{res5}) provided that
\begin{align}\label{quintid}
 &(n-1+G)S_{(n-1)miklj} +(m-1+G)S_{n(m-1)iklj} + (i-1+G)S_{nm(i-1)klj}\\
&\hspace{3cm}\nonumber
 - (k+1)S_{nmi(k+1)lj} - (l+1)S_{nmik(l+1)j} - (j+1)S_{nmikl(j+1)}=0
\end{align} 
for all $(n,m,i,j,k,l)$ satisfying $n+m+i-k-l-j = 1$. The proof is an immediate generalization of the corresponding derivation for the cubic case given in section 3 of \cite{AO}. One simply differentiates (\ref{Zdef}) with respect to $t$ and applies (\ref{res5}).

We note that while the conservation of (\ref{Zdef}) generalizes to the quintic case, we are not aware of a similar generalization of the invariant manifolds and the corresponding analytic solutions mentioned for the cubic case under (\ref{Zdef}). On the other hand, the generating function (\ref{S_sol}) and the stationary solutions (\ref{mode0fam}) and  (\ref{uNder}) do have natural quintic counterparts, as we shall proceed to demonstrate.

\subsection{The generating function}

We would like to convert the condition (\ref{quintid}) into an explicit solution for the generating function
\begin{equation}
S(z_1,z_2,z_3,s_1,s_2,s_3) = \sum_{n_i,k_j=0}^{\infty} S_{n_1n_2n_3k_1k_2k_3} z_1^{n_1} z_2^{n_2}z_3^{n_3}s_1^{k_1}s_2^{k_2}s_3^{k_3}.
\label{quintgendef}
\end{equation}
From (\ref{quintid}), one must have
\beq
\sum_{i=1}^3\left[ z_i^2\del_{z_i}+Gz_i-\del_{s_i}\right]S=0.
\label{Seq1}
\eeq
By the symmetries of $S_{n_1n_2n_3k_1k_2k_3}$, this also implies
\beq
\sum_{i=1}^3\left[ s_i^2\del_{s_i}+Gs_i-\del_{z_i}\right]S=0.
\label{Seq2}
\eeq
These two equations are only compatible if the commutator of the two operators in square brackets annihilates $S$ as well, giving
\beq
\sum_{i=1}^3\left[ z_i\del_{z_i}-s_i\del_{s_i}\right] S=0.
\label{Seq3}
\eeq
Equations (\ref{Seq1}-\ref{Seq3}) can be solved using the method of characteristics, as in \cite{AO}, to yield three alternative representations for $S$:
\begin{eqnarray}
S & = & \frac{A\left(s_1 - \frac{1}{z_1},s_1 - \frac{1}{z_2},s_1 - \frac{1}{z_3},s_1 - s_2, s_1 - s_3\right)}{(z_1z_2z_3)^G}\\
S & = & \frac{B\left(z_1 - z_2, z_1 - z_3, z_1 - \frac{1}{s_1},  z_1 - \frac{1}{s_2},  z_1 - \frac{1}{s_3}\right)}{(s_1s_2s_3)^G}\\
S & = & C(z_1s_1,z_1s_2,z_1s_3,z_2s_1,z_2s_2,z_2s_3,z_3s_1,z_3s_2,z_3s_3),
\end{eqnarray}
where $A$, $B$ and $C$ are arbitrary functions. These can be rewritten as
\begin{eqnarray}
S & = & \frac{\tilde A\left(s_1 - \frac{1}{z_1},s_1 - \frac{1}{z_2},s_1 - \frac{1}{z_3},s_1 - s_2, s_1 - s_3\right)}{\Big[\prod_{i,j=1}^3 (1-z_is_j)\Big]^{G/3}}\label{Stilde1}\\
S & = & \frac{\tilde B\left(z_1 - z_2, z_1 - z_3, z_1 - \frac{1}{s_1},  z_1 - \frac{1}{s_2},  z_1 - \frac{1}{s_3}\right)}{\Big[\prod_{i,j=1}^3 (1-z_is_j)\Big]^{G/3}}\\
S & = &\frac{\tilde C(z_1s_1,z_1s_2,z_1s_3,z_2s_1,z_2s_2,z_2s_3,z_3s_1,z_3s_2,z_3s_3)}{\Big[\prod_{i,j=1}^3 (1-z_is_j)\Big]^{G/3}}.\label{Stilde3}
\end{eqnarray}
One can easily verify that the ratio of $A$ and $\tilde A$ is expressible through the arguments of $A$ (which are the same as the arguments of $\tilde A$), and similarly for $B$ and $C$.

From (\ref{Stilde1}-\ref{Stilde3}), $\tilde A$, $\tilde B$ and $\tilde C$ are the same function, which should be expressible in three different ways through the indicated arguments. Furthermore, $\tilde A=\tilde B=\tilde C$ must have a regular power series expansion around $z_i=s_i=0$ in order to provide a legitimate generating function of the form (\ref{quintgendef}). While the denominator of (\ref{Stilde1}-\ref{Stilde3}) has such an expansion, and for $\tilde C$ one simply needs regularity around the zeros of its arguments, the presence of $1/z_i$ and $1/s_i$ in the arguments of $\tilde A$ and $\tilde B$ requires imposing further constraints on these functions in order for the power series expansion of $S$ to exist. 

Consider $\tilde A$ first. By writing $s_1-1/z_1-(s_1-s_2)=s_2-1/z_1$ and so on, we may express $\tilde A$ as a function of nine arguments $s_i-1/z_k$ for all possible values of $i$ and $k$. Of course, only five of these new arguments are independent, but the advantage is that the expression becomes more symmetric with respect to permutations of $z_i$ and $s_i$. Now, we need ${\tilde A}(s_i-1/z_k)$ to have a regular power series expansion near $z_i=s_i=0$, which corresponds to infinite values of all of its nine arguments. It is hence more convenient to express $\tilde A$ through $-1/(s_i-1/z_k)=z_k/(1-z_ks_i)$ so that $z_i=s_i=0$ corresponds to zero values of the new arguments. Since we know from (\ref{Stilde3}) that $\tilde A$ must be expressible through $z_is_j$ only,  it must depend on the ratios of $z_k/(1-z_ks_i)$ and  $z_k/(1-z_ks_j)$, in which $z_k$ cancels out. Thus, $\tilde A$ is a function of the ratios $(1-z_ks_i)/(1-z_ks_j)$. By a similar argument, $\tilde B$ is a function of the ratios $(1-z_is_k)/(1-z_js_k)$.
The combinations of $z_i$ and $s_i$  that are expressible both through the arguments of $\tilde A$ and $\tilde B$ are
\beq
u_{ijkl}=\frac{(1-z_is_k)(1-z_js_l)}{(1-z_is_l)(1-z_js_k)}.
\label{crs}
\eeq
These can be recognized as (a subset of)  the conformally invariant cross-ratios \cite{appconf} of the six points $(z_1,z_2,z_3,1/s_1,1/s_2,1/s_3)$ on the complex plane. The generating function takes the form
\beq
S(z_i,s_i)=\frac{{\cal F}(\ln (u_{ijkl}))}{\Big[\prod_{n,m=1}^3 (1-z_ns_m)\Big]^{G/3}},
\label{genquint}
\eeq
where $\cal F$ is an arbitrary function, and the logarithm is introduced in the definition to make it easier to account for the symmetries of $S$, as permutations of $z_i$ and $s_i$ act particularly straightforwardly on the logarithms of cross-ratios (\ref{crs}). The symmetries of $S$, of course, translate into a set of symmetries of $\cal F$, but we will not need to characterize them explicitly for our present purposes.
 
\subsection{Complex plane representation and stationary states}

With the generating functions (\ref{genquint}), the considerations of sections \ref{sec:cpl} and \ref{sec:stst} directly generalize to the quintic case. One first introduces $\beta_n$ as in (\ref{betaSdef}) and the corresponding generating functions $u$ and $\tilde u$ as in (\ref{udef}-\ref{betaint}). Then, (\ref{res5}) is converted to the corresponding equation in terms of the generating functions, analogous to (\ref{complexeq}):
\begin{align}
&\frac{i}{\Gamma(G)}\del_t\del_z^{G-1}(z^{G-1} u(t,z))=
\frac1{(2\pi i)^5}\oint\frac{dz_2}{z_2} \oint\frac{dz_3}{z_3}\oint\frac{ds_1}{s_1} \oint\frac{ds_2}{s_2} \oint\frac{ds_3}{s_3} \label{complexeq5}  \\
&\hspace{3cm}\times S(z,z_2,z_3,1/s_1,1/s_2,1/s_3)\,\, \tilde u(t,z_2)\tilde u(t,z_3) u(t,s_1) u(t,s_2) u(t,s_3),\nonumber
\end{align}
with $S$ given by (\ref{genquint}).

As the generating function (\ref{genquint}) is very similar to (\ref{S_sol}), the structure of equation (\ref{complexeq5}) is directly parallel to (\ref{complexeq}) and (\ref{complexeq2}). In particular, there are solutions in the form of stationary states
\beq
u(t,z)=e^{-i\lambda t}u(z),\qquad\del_z^{G-1}(z^{G-1} u(z))=\frac{(\bar p -z)^N}{(1-pz)^{N+G}}
\label{ststquint}
\eeq
for any nonnegative integer $N$ and any complex number $p$. The proof simply retraces the steps of section \ref{sec:stst}. Note that because $\cal F$ in (\ref{genquint}) only depends on the conformal cross-ratios, there will be a direct analog of the differentiation formula (\ref{diffG}), which plays a key role in the derivation.

\subsection{$G\to\infty$ limit}

We have already remarked in section \ref{sec:cpl} that there is a special $G\to\infty$ limit of our construction for the cubic case, which has been discussed in detail in \cite{AO}. A similar limit exists for the quintic generalization of the current section. Namely, instead of (\ref{fdef}), one defines
\beq
f^\infty_n=\frac1{\sqrt{n!}},
\eeq
which is used instead of $f_n$ to define $\beta_n$ as in (\ref{betaSdef}) and $S$ as in (\ref{Squintdef}),
\beq
S_{n_1n_2n_3k_1k_2k_3}=\frac{C_{n_1n_2n_3k_1k_2k_3}}{\sqrt{n_1!\,n_2!\,n_3!\,k_1!\,k_2!\,k_3!}}.
\label{Squintinf}
\eeq
 Then, instead of (\ref{quintid}), one imposes
\begin{align}\label{idinf}
 &S_{(n-1)miklj} +S_{n(m-1)iklj} + S_{nm(i-1)klj}\\
&\hspace{3cm}\nonumber
 - (k+1)S_{nmi(k+1)lj} - (l+1)S_{nmik(l+1)j} - (j+1)S_{nmikl(j+1)}=0
\end{align} 
for all sets of indices satisfying $n+m+i=k+l+j+1$,
which guarantees the conservation of 
\beq
Z  =  \sum_{n=0}^{\infty}\sqrt{n+1}\,\bar{\alpha}_{n+1} \alpha_n,
\label{Zinf}
\eeq
and the presence of associated symmetries.

One could pursue a construction of stationary states bifurcating from individual modes for this case, as in section \ref{sec:stst}, but there is a shortcut available that makes it unnecessary. For the special $G\to\infty$ limit, unlike for general values of G, the finite form of symmetry transformations generated by (\ref{Zinf}) is known explicitly. Such transformations have appeared under the name of magnetic translations in the literature on the Lowest Landau Level approximation for Bose-Einstein condensates \cite{GHT}. In terms of the generating function
\beq
u(t,z)=\sum_{n=0}^\infty \frac{\alpha_n(t)\,z^n}{\sqrt{n!}},
\eeq
applying the transformation
\beq
u(t,z)\mapsto u(t,z-\bar p)\,e^{pz-|p|^2/2}
\label{inftrans}
\eeq
maps solutions of (\ref{res5}) with the interaction coefficients satisfying (\ref{Squintinf}-\ref{idinf}) into solutions. Applying these transformation to the generating functions of single-mode solutions, $u\sim z^N$, gives the $G\to\infty$ analogs of the stationary states (\ref{ststquint}).


\section{Quintic PDEs with partially solvable resonant systems}

We shall now present two examples of quintic PDEs of mathematical physics whose resonant systems fall in the partially solvable class we have outlined and benefit from the structures developed in our derivations. The first example is the quintic one-dimensional nonlinear Schr\"odinger equation in a harmonic trap. The resonant system for this equation has been previously studied in mathematical literature \cite{fennell}, while motivations to consider quintic nonlinearities from the standpoint of Bose-Einstein condensate physics are given in \cite{NLSquint}. Our purpose is to clarify how this case fits in our framework. The second example, which is the resonant system of the quintic conformally invariant wave equation on a two-sphere, is novel. (Additionally, in the appendix, we give a few extra quintic resonant systems that fall into our special class, obtained by directly generalizing known cubic examples, rather than deriving them from concrete quintic PDEs.)

\subsection{Quintic nonlinear Schr\"odinger equation on R$^1$}

Mathematically rigorous considerations of the resonant approximation to the quintic one-dimensional nonlinear Schr\"odinger equation in a harmonic trap are given in \cite{fennell}. In a nutshell, one starts with the quintic analog of the nonlinear Schr\"odinger equation (\ref{NLS1d}) given by
\begin{equation}
i\,\frac{\partial \Psi}{\partial t}=\frac12\left(-\frac{\partial^2}{\partial x^2}+x^2\right)\Psi +g|\Psi|^4\Psi.
\label{NLS1d5}
\end{equation}
and performs manupulations identical to (\ref{NLS1dlin}-\ref{NLSpreres}) to obtain a resonant system of the form (\ref{res5}) with the interaction coefficients given by
\begin{equation}
C_{nmiklj} = \frac{1}{2^{n+m+i}\sqrt{n!m!i!k!l!j!}}\int_{-\infty}^{\infty} dx e^{-3 x^2} H_n H_m H_i H_k H_l H_j.
\label{intcoeffNLS1}
\end{equation}
Here, $H_n$ are Hermite polynomials. These interaction coefficients satisfy  (\ref{Squintinf}-\ref{idinf}), and thus the resonant system belongs to the special $G\to\infty$ limit of our class.

To prove that (\ref{intcoeffNLS1}) satisfies (\ref{Squintinf}-\ref{idinf}), recall the following identities for the Hermite polynomials 
\begin{align}
& H_{n+1} = 2x H_n - \partial_x H_n, \label{eq:_id_1_H}\\
&\partial_x H_n = 2n H_{n-1}, \label{eq:_id_2_H}\\
&H_{n+1} = 2x H_n - 2n H_{n-1}. \label{eq:_id_3_H}
\end{align}
Omitting the irrelevant numerical prefactor and remembering that $n+m+i=k+l+j+1$, we obtain for the left-hand side of (\ref{idinf})
\begin{equation*}
\int_{-\infty}^{\infty} dx e^{-3 x^2} [ 2nH_{n-1}H_m H_i H_k H_l H_j + 2m H_{n}H_{m-1} H_i H_k H_l H_j + 2i H_{n}H_m H_{i-1} H_k H_l H_j -
\end{equation*}
\begin{equation*}
- H_n H_mH_i H_{k+1}H_lH_j - H_n H_mH_i H_{k}H_{l+1}H_j - H_n H_mH_i H_{k}H_lH_{j+1}] 
\end{equation*}
\begin{equation*}
=\int_{-\infty}^{\infty} dx e^{-3 x^2} [ \partial_x \left(H_nH_mH_i\right) H_kH_lH_j -  6x H_n H_mH_i   H_lH_j +  H_n H_mH_i \partial_x\left(H_{k}H_lH_{j}\right)] 
\end{equation*}
\begin{equation*}
 = \int_{-\infty}^{\infty} dx e^{-3 x^2} [ \partial_x \left(H_nH_mH_iH_kH_lH_j\right)  -  6x H_n H_mH_i   H_lH_j ] 
\end{equation*}
\begin{equation*}
=\int_{-\infty}^{\infty} dx  \partial_x[e^{-3 x^2} H_nH_mH_iH_kH_lH_j ] 
= e^{-3 x^2} H_n(x)H_m(x)H_i(x)H_k(x)H_l(x)H_j(x)  \Big{|}_{-\infty}^{\infty} = 0.
\end{equation*}
Thus, (\ref{idinf}) is satisfied and (\ref{Zinf}) is conserved. Stationary solutions can be straightforwardly constructed using the transformation (\ref{inftrans}).


\subsection{Conformally invariant quintic wave equation on S$^2$}

We now turn to the conformally invariant quintic wave equation on a two-sphere, where the  structures we display are novel. A cubic precursor of these structures is the conformal flow originating from the cubic wave equation on a three-sphere and studied in \cite{CF,BHP}. One of the questions that has triggered our present study of quintic nonlinearities is whether the properties of the conformal flow generalize to the quintic case. Existence of such generalization is by no means guaranteed.

Consider the (2+1)-dimensional Einstein cylinder $\mathbb{R}\times \mathbb{S}^2$ with the  metric
\begin{equation}\label{metric}
   g= -dt^2 + d\vartheta^2+\sin^2{\vartheta} d\varphi^2
  \end{equation}
and put on it a real scalar field $\phi$ satisfying
 the conformally invariant quintic wave equation\footnote{Note that solutions to (\ref{eq-conf}) satisfying $\phi=0$ at the equator of the two-sphere can be mapped by a standard conformal transformation to solutions for a conformally invariant quintic wave equation on the three-dimensional Anti-de Sitter spacetime with the Dirichlet boundary conditions at infinity. This map connects our equation to topics of the much-studied AdS$_3$/CFT$_2$-correspondence.}
 \begin{equation}\label{eq-conf}
  \left(\square_g -\frac{1}{8} R(g)\right) \phi -\phi^5 =0\,,
  \end{equation}
  where $\square_g:=g^{\mu\nu}\nabla_{\mu}\nabla_{\nu}$ and $R(g)$ are the wave operator and the Ricci scalar associated with $g$. We introduce $x=\cos{\vartheta}$ and impose rotational symmetry by assuming that $\phi=f(t,x)$. Then, equation \eqref{eq-conf} reduces to
  \begin{equation}\label{eq}
  f_{tt}-\partial_x \left((1-x^2) f_x\right)  + \frac{1}{4} f  + f^5 =0\,.
  \end{equation}
Decomposing small (weakly nonlinear) solutions with amplitudes of order $\eps$ into Legendre polynomials satisfying the standard normalization condition $\int_{-1}^1 P_n\,P_m\,dx=2\de_{nm}/(2n+1)$ as
\beq
f(t,x)=\eps \sum\limits_{n=0}^{\infty} c_n(t) P_n(x),
\eeq
we obtain from \eqref{eq} an infinite system of coupled oscillators
\begin{equation}\label{fourier}
 \frac{d^2 c_n}{dt^2} + \omega_n^2 c_n = -\om_n\eps^4\,\sum\limits_{jklmi} C_{njklmi}\, c_j c_k c_l c_m c_i,
\end{equation}
where $\omega_n=n+\frac{1}{2}$ and
\begin{equation}\label{C}
 C_{njklmi}=\int_{-1}^{1} P_n(x) P_j(x) P_k(x) P_l(x) P_m(x) P_i(x) \,dx.
\end{equation}

We now develop a resonant approximation to (\ref{fourier}) at small $\eps$, based on the standard time-averaging techniques \cite{murdock}. One first introduces the complex normal mode amplitudes $\alpha_n$, which will be our new dynamical variables
\beq
c_n(t)=\alpha_n(t) e^{-i\om_n t}+ \bar\alpha_n(t) e^{i\om_n t},\qquad  \frac{d c_n(t)}{dt}=-i\om_n\left(\alpha_n(t) e^{-i\om_n t}- \bar\alpha_n(t) e^{i\om_n t}\right).
\eeq
Substituting this into (\ref{fourier}), one obtains schematically
\beq
i\om_n \frac{d\alpha_n}{dt}=\om_n\eps^4\sum C_{njklmi}\, \alqu_j \alqu_k \alqu_l \alqu_m \alqu_i\, e^{i\Omega_{njklmi}t},
\label{CF5preres}
\eeq
where $\alqu_n$ may be either $\alpha_n$ or $\bar\alpha_n$ (all such choices must be summed over) and 
\beq
\Omega_{njklmi}=\om_n\pm \om_j\pm \om_k\pm \om_l\pm \om_m\pm \om_i,
\label{Enjklmi}
\eeq
where the plus signs are chosen if the corresponding $\alqu_n$ occurs as $\bar\alpha_n$, and the minus signs if it occurs as  $\alpha_n$.

By the standard lore of time-averaging \cite{murdock}, at small $\eps$, nonresonant interactions corresponding to nonzero $\Omega_{njklmi}$ may be dropped from (\ref{CF5preres}). Keeping only the resonant couplings satisfying $\Omega_{njklmi}=0$ results in an accurate approximation on time scales $1/\eps^4$ for small $\eps$, and amounts to implementing the resonant approximation.

Before we state the final form of our quintic resonant system, it is important to point out an additional simplification that occurs for our specific case. There are many possible choices of signs in (\ref{Enjklmi}), but it turns out that the interaction coefficients (\ref{C}) vanish unless there are exactly three plus signs and three minus signs total. This is analogous to the selection rules that have been extensively studied in the context of resonant systems in Anti-de Sitter spacetime \cite{CEV,Yang,EN}. We now sketch a proof of this claim.

Consider first the case where (\ref{Enjklmi}) has one plus sign and five minus signs. Then, the resonant condition is $\Omega_{njklmi}=\om_n-\om_j- \om_k- \om_l- \om_m- \om_i=0$, which means that $n=j+k+l+m+i+2$. But then the degree of $P_n$ in (\ref{C}) is higher than the total degree of the polynomial $P_j P_k P_lP_mP_i$, and hence the corresponding $C$ vanishes by orthogonality of the Legendre polynomials and does not contribute to (\ref{CF5preres}). Consider now the case where (\ref{Enjklmi}) has two plus signs and four minus signs. By the index permutation symmetry of $C$, the two plus signs can be associated with the indices $n$ and $j$, so that the resonant condition is $\Omega_{njklmi}=\om_n+\om_j- \om_k- \om_l- \om_m- \om_i=0$, which means $n+j=k+l+m+i+1$. But the Legendre polynomials satisfy $P_n(-x)=(-1)^nP_n(x)$. Hence, $P_nP_j P_k P_lP_mP_i$ is reflection-odd, and its integral from $-1$ to $1$ in (\ref{C}) is zero. The cases with four and five plus signs reduce to the previous two cases after renaming the indices. Thus, the only resonant condition that may contribute to the resonant approximation to (\ref{CF5preres}) is (\ref{Enjklmi}) with three plus and three minus signs (which corresponds to two $\bar\alpha$'s and three $\alpha$'s), as claimed.

Putting everything together and absorbing $\eps^4$ together with any numerical factors into a redefinition of time, we arrive at the following resonant approximation to (\ref{fourier}), which we call the {\em quintic conformal flow}:
\begin{equation} \label{flow}
i \frac{d\alpha_n}{dt}=\hspace{-3mm} \sum\limits_{n+n_2+n_3=n_4+n_5+n_6}\hspace{-3mm} C_{n n_2 n_3 n_4 n_5 n_6} \bar\alpha_{n_2} \bar \alpha_{n_3} \alpha_{n_4} \alpha_{n_5} \alpha_{n_6}.
\end{equation}
The interactions coefficients can be expressed as
\begin{equation}
 C_{n_1 n_2 n_3 n_4 n_5 n_6} = \frac{1}{1+N}\,\sum_{j_1=0}^{n_1} \sum_{j_2=0}^{n_2} \sum_{j_3=0}^{n_3} \sum_{j_4=0}^{n_4} \sum_{j_5=0}^{n_5} \sum_{j_6=0}^{n_6} (-1)^J \frac{\prod_{k=1}^6 {\binom{n_k}{j_k}}^2}{\binom{N}{J}},
\end{equation}
where $N=\sum_{k=1}^6 n_k$ and  $J=\sum_{k=1}^6 j_k$, though we shall not use this explicit formula. The corresponding analysis together with a combinatorial interpretation of the integrals (\ref{C}) can be found in \cite{GJZ}.

We shall now prove that the interaction coefficients (\ref{C}) satisfy (\ref{quintid}) with $G = 1$, and thus (\ref{flow}) belongs to the class of resonant systems analyzed in our paper. We first notice the identities
\begin{align}
&n P_{n-1} = (2n+1) x P_n - (n+1) P_{n+1}, \label{eq:_id_1}\\
&(x^2-1)\partial_x P_n = n x P_n - n P_{n-1}, \label{eq:_id_2}\\
&(x^2-1)\partial_x P_n + (n+1) x P_n = (n+1)P_{n+1}. \label{eq:_id_3}
\end{align} 
From (\ref{eq:_id_1}),
\begin{equation*}
n P_{n-1}P_m P_i P_k P_l P_j + m P_nP_{m-1} P_i P_k P_l P_j + i P_nP_m P_{i-1} P_k P_l P_j =  (2(n+m+i) + 3) x P_n P_m P_i P_k P_l P_j 
\end{equation*}
\begin{equation*}
- (n+1) P_{n+1}P_m P_i P_k P_l P_j  - (m+1)P_nP_{m+1} P_i P_k P_l P_j - (i+1) P_n P_m P_{i+1} P_k P_l P_j .
\end{equation*}
Hence, the left-hand side of (\ref{quintid}) becomes
\begin{equation*}
\int_{-1}^{1}dx\ [ (2(n+m+i) + 3) x P_n P_m P_i P_k P_l P_j - (n+1) P_{n+1}P_m P_i P_k P_l P_j  - (m+1)P_nP_{m+1} P_i P_k P_l P_j
\end{equation*}
\begin{equation*}
 - (i+1) P_n P_m P_{i+1} P_k P_l P_j  - (k+1)P_n P_m P_i P_{k+1} P_l P_j - (l+1)P_n P_m P_i P_k P_{l+1} P_j - (j+1)P_n P_m P_i P_k P_l P_{j+1}].
\end{equation*}
Using (\ref{eq:_id_3}) on the terms with minus signs, this is written as
\begin{align*}
 &\int_{-1}^{1}dx\  [ (2(n+m+i) + 3) x P_n P_m P_i P_k P_l P_j - (x^2-1)\partial_x \left(P_nP_mP_iP_kP_lP_j\right) \\
&
\hspace{5cm}- (n+m+i+k+l+j + 6) x P_nP_mP_iP_kP_lP_j].
\end{align*}
Then, remembering that $n+m+i = k + l+j + 1$, one gets
\begin{equation*}
\int_{-1}^{1}dx\left(-2x - (x^2-1)\partial_x \right)P_n P_m P_i P_k P_l P_j =\int_{-1}^{1}dx \partial_x \left((x^2-1) P_n P_m P_i P_k P_l P_j \right) =
\end{equation*}
\begin{equation*}
= (x^2-1) P_n (x)P_m(x)P_i(x) P_k(x) P_l(x) P_j(x) \Big{|}_{-1}^{1} = 0.
\end{equation*}
Hence, (\ref{quintid}) is satisfied, and the resonant system (\ref{flow}) benefits from the conserved quantity (\ref{Zdef}) and the stationary solutions (\ref{ststquint}) with $G=1$.


\section{Outlook}

We have explored properties of cubic resonant systems of the form (\ref{ressyst}) with the interaction coefficients satisfying (\ref{AOdef}), and quintic resonant systems of the form (\ref{res5}) with the interaction coefficients satisfying (\ref{quintid}). By constructing complex plane representations (\ref{complexeq}), (\ref{complexeq2}) and (\ref{complexeq5}), we were able to establish families of stationary states (\ref{uNder}) bifurcating from every individual mode. Our results have direct implications for a number of equations of mathematical physics whose resonant systems fall in our class, in relation to Bose-Einstein condensates \cite{BBCE}, the Schr\"odinger-Newton system in a harmonic potential \cite{BEF}, and relativistic wave equations in highly symmetric spacetimes \cite{CF,BEL}. Examples with quintic nonlinearities include the nonlinear Schr\"odinger equation in a one-dimensional harmonic trap, previously treated from a mathematical perspective in \cite{fennell}, and the conformally invariant quintic wave equation on a two-sphere, brought forth in a our present study. We have also presented in the appendix a few extra quintic resonant systems that benefit from the analytic stuctures we have formulated, which have been obtained by a generalization of explicitly known cubic resonant systems. Our formalism with its defining conditions  (\ref{AOdef}) and (\ref{quintid}), as well as the examples from the appendix, furthermore admit a natural extension to higher order nonlinearities (which we do not explicitly pursue).

The class of resonant systems we have studied includes known representatives originating as resonant approximations \cite{CF,BEL,BEF} to nonlinear dynamics in AdS spacetimes. The latter topic is of appreciable significance in the area of AdS/CFT correspondence, where such studies connect to the physics of thermalization in the dual field theories \cite{therm,relax}. While the dynamical equations in \cite{CF,BEL,BEF} do not always include gravitational backreaction on the AdS metric, such nonlinear probe fields have also been studied in the context of AdS/CFT correspondence, see, e.g., \cite{Das}.

We believe that the complex plane representations for our resonant systems hold much more power than we have explicitly displayed in our treatment. As we mentioned in the introduction, the complex plane representation for the conformal flow \cite{CF} can be used to give an elegant proof \cite{gerard_quintic} to the rather nontrivial fact that there are stationary states with generating functions in the form of arbitrary Blaschke products
\beq
u(t,z)=e^{-i\lambda t}\,\frac{(\bar p_1-z)\cdots(\bar p_k-z)}{(1-p_1z)\cdots (1-p_kz)}
\eeq
for any set of complex numbers $(p_1,\ldots,p_k)$ and any $k$. Similarly, a complex plane represention for the LLL equation has been used in \cite{GGT} to obtain powerful results on classification of stationary states and properties of their zeros (the latter subject being of pivotal importance in the physics of Bose-Einstein condensates). It remains to be seen what further conclusions may emerge from the analysis of our complex plane representations, both in general, and in application to specific physically motivated representatives in our classes of resonant systems.


\section*{Acknowledgments}

This research has been supported by FPA2014-52218-P and FPA2017-84436-P from Ministerio de Economia y Competitividad, by Xunta de Galicia ED431C 2017/07, by European Regional Development Fund (FEDER), by Grant Mar\'ia de Maetzu Unit of Excellence MDM-2016-0692, by Polish National Science Centre grant number 2017/26/A/ST2/00530 and by CUniverse research promotion project by Chulalongkorn University (grant CUAASC). A.B. thanks the Spanish program ``ayudas para contratos predoctorales para la formaci\'on de doctores 2015'' and its mobility program for his stay at Jagiellonian University, where part of this project was developed.


\section*{Appendix: Additional examples of quintic partially solvable\\\rule{2.7cm}{0cm} resonant systems}

We list a few quintic systems satisfying (\ref{quintid}) or (\ref{idinf}) obtained by a direct generalization of known cubic systems satisfying (\ref{AOdef}):

\begin{itemize}
	\item  The interaction coefficients 
		 \begin{equation}
		 	S_{nmiklj} = \frac{1}{(n+m+i +1)(n+m+i + 2)}
		 \end{equation}
	  satisfy (\ref{quintid}) with $G = 1$. This is a generalization of the maximally rotating cubic resonant system on a three-sphere from \cite{BEL}.

	\item  The interaction coefficients 
		 \begin{equation}
		 	S_{nmiklj} = \frac{\Gamma(n+\delta)\Gamma(m+\delta)\Gamma(i+\delta)\Gamma(k+\delta)\Gamma(l+\delta)\Gamma(j+\delta)\Gamma(n+m+i+1)}{\Gamma(n+1)\Gamma(m+1)\Gamma(i+1)\Gamma(k+1)\Gamma(l+1)\Gamma(j+1)\Gamma(n+m+i+3\delta)}
		 \end{equation}
	  satisfy (\ref{quintid}) with  $G = \delta$, which can be an arbitrary positive real number.  This is a generalization of the maximally rotating cubic resonant systems on Anti-de Sitter spacetimes from \cite{BEL}.
		
		 \item The interaction coefficients 
		 \begin{equation}
		 	S_{nmiklj} = \frac{8}{\pi}\int_{0}^{\pi} \frac{dx}{\sin^2 x}\sin(n+1)x\sin(m+1)x\sin(i+1)x\sin(k+1)x\sin(l+1)x\sin(j+1)x
		 \end{equation}
	 satisfy (\ref{quintid}) with $G = 2$. The cubic prototype is
		 \begin{equation}
		 	S_{nmkl} = \frac{2}{\pi}\int_{0}^{\pi} \frac{dx}{\sin^2 x}\sin(n+1)x\sin(m+1)x\sin(k+1)x\sin(l+1)x = \text{min}(n,m,k,l)+1,
		 \end{equation}
which is the conformal flow \cite{CF}.

	 \item The interaction  coefficients
		 \begin{equation}
		 	S_{nmiklj} = \frac{1}{3^{n+m+i}} \frac{(n+m+i)!}{n!m!i!k!l!j!}
		 \end{equation}
	 Satisfy (\ref{idinf}) and thus correspond to the $G\to\infty$ limit in our class of systems. The cubic prototype is the LLL equation \cite{GHT,GT,BBCE,GGT}.
		 
\end{itemize}


\end{document}